\documentclass[12pt,english,twoside, reqno]{article}

\usepackage{graphicx}
\usepackage{amssymb}
\usepackage{textcomp}

\usepackage{cite}

\usepackage{bm}

\usepackage{mathabx}

\usepackage{dingbat}

\usepackage{pifont}

\usepackage{fancyhdr}
\usepackage[english,french]{babel}
\usepackage[latin1]{inputenc}
\usepackage{tipa}
\usepackage{bbm}

\usepackage{graphicx}
\usepackage{amssymb}
\usepackage{textcomp}
\usepackage{stmaryrd}

\begin{document}

\newcommand{\nin}{\noindent}
\newcommand{\vv}{\vskip 0.25 cm}
\newcommand{\vvv}{\vskip 0.3 cm}
\newcommand{\vvvv}{\vskip 0.5 cm}

\newcommand{\beq}{\begin{equation}}
\newcommand{\enq}{\end{equation}}
\newcommand{\und}{\underline}
\newcommand{\biz}{\begin{itemize}}
\newcommand{\eiz}{\end{itemize}}
\newcommand{\di}{\displaystyle}
\newcommand{\bc}{\begin{center}}
\newcommand{\ec}{\end{center}}
\newcommand{\uta}{\begin{array}[t]{c}
{A'} \\
\stackrel{\sim}{\stackrel{~~}{~~}}
\end{array}}
\newcommand{\utx}{\begin{array}[t]{c}
{\xi} \\
\stackrel{\sim}{\stackrel{~~}{~~}}
\end{array}}
\newcommand{\pa}{\partial}
\newcommand{\g}{\gamma}
\newcommand{\ep}{\epsilon}
\newcommand{\cp}{\stackrel{\circ}{p}}
\newcommand{\ck}{\stackrel{\circ}{k}}

\newcommand{\uz}{\stackrel{\circ}{U}}
\newcommand{\az}{\stackrel{\circ}{A}}
\newcommand{\bz}{\stackrel{\circ}{B}}
\newcommand{\vz}{\stackrel{\circ}{V}}
\newcommand{\Vec}{\stackrel{\longrightarrow}}
\newcommand{\ddd}{\stackrel{{...}}}

\newcommand{\ov}{\overline}

\newcommand{\vp}{\varphi}

\newcommand{\bms}{\boldsymbol}

%{\sf

\vskip 1.5 true cm

\centerline{\bf \large Classical dynamical symmetries and geometry of trajectories}

\vskip 0.5 true cm 

\vskip 0.75 cm
\centerline{{\bf Christian CARIMALO}\footnote{{\bf christian.carimalo@sorbonne-universite.fr}}}

\vskip 0.75 cm
\centerline{Sorbonne Universit\'e, Campus Pierre et Marie Curie}
\centerline{4 Place Jussieu, 75005 Paris, France}

\vskip 1.5cm
\centerline{ABSTRACT}

\vv \nin It is shown that all spherical symmetric potentials are capable of producing dynamical symmetries in classical one-body motions, thanks to the inevitable existence of symmetry axes associated with turning points for corresponding trajectories. This will definitely expand the class of maximally superintegrable one-body motions in central potentials that until now was considered to include only the Newtonian and Hookean cases. A simple method is proposed to identify and characterize these dynamical symmetries.  

\vskip 0.75 cm

 
\vskip 0.75cm

\section{\large Introduction} 

\vv \nin In classical and non-relativistic mechanics, the one-body motion in a spherical symmetric potential is certainly a well known and long studied problem. However, in a general case, the question of whether such system has an additional dynamical (or ``hidden") symmetry  has remained unanswered until now. This is because of ignorance of what might be the true origin of such a symmetry and lack of a simple method to reveal and describe it. In this regard, the general method of Lie's transformations of differential equations is disapointing as it leads either to intractable equations when using contact transformations or to a well known result (mechanical similarity) when using point transformations : see Ref. \cite{cari1}. Discovering a dynamical symmetry could therefore appear as a matter of luck. Yet, the two famous examples of the Newtonian and the Hookean systems suggest a strong and simple correlation between the appearance of an internal symmetry and the geometrical properties of the trajectories. For the Newtonian system with the attractive interaction potential $V(r) = - K/r$ ($K>0$), its dynamical symmetry is characterized by the celebrated Laplace-Runge-Lenz (LRL) vector, Refs. \cite{LAP,RUN,LEN}, which is found to be along a symmetry axis for each trajectory. Together with the angular momentum, this vector generates a group of transformation of trajectories of the same species, currently identified as $SO(4)$ for ellipses or $SO(3,1)$ for hyperbolas. For the Hookean system with the harmonic potential $V(r) = K^\prime r^2/2$ ($K^\prime >0$), its dynamical symmetry is characterized by one of the two vectors defining the symmetry axes of trajectories, and that vector also takes part in a group of transformation of the corresponding trajectories (ellipses) :  see Ref. \cite{car2}. Taking this correlation as a guiding idea, we show below how it works as well for any other spherical symmetric potential and we develop a simple method to identify and characterize the resulting dynamical symmetry.

\vv \nin The correlation between dynamical symmetries and geometrical properties of trajectories for central potentials has already been considered by various authors in the past, Ref. \cite{BRS, FRD,MUK}. These authors can be considered as precursors of the subject although they treated it in a rather vague way. The authors of Ref. \cite{BRS} gave some importance to a characteristic point of trajectories from which a symmetry could emerge,  such as the perihelion for ellipses, but added in a footnote that ``Every point of a trajectory can be considered as a characteristic point, except of course if the trajectory is a circle", while the author of Ref. \cite{FRD} said that ``$O_4$ and $SU_3$ are dynamical symmetries for a central potential essentially because motion takes place in a fixed plane". In fact, all these articles have mainly focused on mathematical aspects of symmetries from the point of view of group theory (with possible extension to quantum mechanics), and none of them has emphasized what justifies the very existence of a symmetry group, namely its role with respect to trajectories which consists in a continuous mapping of trajectories of a same species, see Ref, \cite{cari1}.

% ==============================================
\section{\large Main known facts about trajectories }

\setcounter{equation}{0}
\renewcommand{\theequation}{\mbox{2.}\arabic{equation}}

\vv \nin The motion considered here is that of a particle $P$ in an inertial frame having as origin  the source $O$ of a spherical symmetric potential acting on $P$.  We will use the following notations : ${\bms r}$ is the position vector of $P$ relative to $O$, and $r = ||{\bms r}||$ ; ${\bms  v}$ and ${\bms p}= m{\bms v}$ are its velocity and momentum vectors, respectively, and $v = ||{\bms v}||$, $p=||{\bms p}|| = mv$ ; finally, implicit summation on repeated indices is also assumed.

\vv \nin Let us recall some known facts and formulas about trajectories.

\vv \nin Thanks to the spherical symmetry of the potential, the angular momentum of the particle, ${\bms L} = {\bms r} \times {\bms p}$, is a conserved vector, and consequently its  trajectory corresponding to a given value of ${\bms L}$ lies in a plane perpendicular to the latter and containing $O$. As the orientation of axes can be freely chosen, $Oz$ is taken along ${\bms L}$, the plane of motion being then the plane $(x, y)$. The position of $P$ in that plane may be defined as usual by the distance $r =OP$ and the polar angle $\theta$ of ${\bms r}$ relative to the axis $Ox$. Since $z=0$, $p_z=0$, only the component $L_z$ of the angular momentum is non-zero, and we have  

$$ L_z =  ||{\bms L}|| = L= m r^2 \dot{\theta} = {\rm const} \label{aires} $$

\vv \nin from which the Kepler's second law (law of equal areas) is deduced. In the following, we exclude the case $L=0$ corresponding to motions along straight lines passing through $O$. The potential being independent of time, the Hamiltonian (or energy) of $P$, $ H = \di{ p^2 \over {2m}} + V(r)$, is also a conserved quantity $E$. The radial and orthoradial components of ${\bms p}$ being respectively $p_r = m \dot{r}$ and $p_\theta = m r \dot{\theta} = L/r$, we reexpress this Hamiltonien as 

\beq H = \di{{m {\dot{r}}^2}\over 2} + U(r),~~{\rm where}~~U(r) = \di{L^2\over{2m r^2}} + V(r) \label{hamiltonien} \enq 

\vv \nin is the effective potential. Writing 
$dH/dt =0$, the distance $r$ is found to vary in time according to the equation 

\beq  \dot{p}_r = m \ddot{r} = - \di{{dU}\over{dr}} \label{eqr} \enq 

 \vv \nin From Eq. (\ref{hamiltonien}), permissible motions of $P$ satisfy 

$$ \di{p^2_r \over{2m}} = E-U(r) \geq 0 $$ 

\vv \nin Turning points are those for which $p_r =0$ or $U(r) =E$. Such points are inevitably  present in closed orbits but may also exist for non-closed and unbounded trajectories. At these points, $p_r$ changes sign. The modulus and the polar components of the velocity are  

$$ v = \sqrt{2m \left(E - V(r) \right)},~~v_\theta = L/(m r),~$$
\beq  v_r =\pm \sqrt{2m(E -V(r)) - L^2/(mr^2)}  \label{vitesse} \enq 

\vv \nin the sign of $v_r$ changing at turning points only. The unit vector of the tangent, ${\bms T} = {\bms v}/v$, has polar components $T_r = v_r/v$, $T_\theta = v_\theta/v = L/(m r v)$. According to Eq. (\ref{vitesse}), at a given $r$, only the component $T_r$ may be different by a simple change of sign. Let us then suppose that the curve representing the trajectory intersects itself at some point. This might occur for closed orbits, when the particle  turns many times around $O$. As the tangent can have only two orientations for a given $r$, this point is a double point (a node). Note also that $v_r$ is linked to the derivative of the polar equation $r(\theta)$ since  

\beq \di{{dr}\over{d\theta}} = \dot{r}/\dot{\theta} = \di{{mr^2 \dot{r}}\over L} \label{polareq} \enq 

\vv \nin and that turning points satisfy as well $\di{{d r}\over{d \theta}}=0$. If a trajectory is symmetrical relatively to an axis passing through $O$ and defined by $\theta =\theta_0$, we must have $r(\theta_0 + \alpha) = r(\theta_0 -\alpha)$ which means that $ \di{{d r}\over{d \theta}}(\theta_0) =0$ (assuming $r(\theta_0)$ well defined) : the point of the trajectory defined by $\theta=\theta_0$ and $r = r(\theta_0)$ is a turning point. 

\vv \nin Using Eq. (\ref{polareq}), the polar angle of the radius vector ${\bms r}$ relatively to some initial position ${\bms {OP}}_0 = r_0 \,{\bms e}_{r0}$ is expressed as

\beq \theta = \theta(r) = \theta_0 + \sqrt{\di{L^2 \over{2m}}} \, \di{\int^r_{r_0}} \di{{ \pm dR}\over{R^2 \sqrt{E-U(R)}}} \label{teta}  \enq   

\vv \nin where the sign of the integrand should be chosen such that $\dot{\theta} = L/(m r^2)$ is always strictly positive : during its motion, the particle $P$ tends to rotate around $O$ in the same direction of rotation. The complete curve of a closed trajectory necessarily surrounds the source $O$. Such a trajectory is bounded and lies inside a circular crown defined by $r_m \leq r \leq r_M$. Points $P_m$ and $P_M$ where $r$ takes its respective extreme values $r_m$ and $r_M$ are turning points. It can be shown that ${\bms OP_m}$ and ${\bms OP_M}$ define symmetry axes of the trajectory : see Appendix A. This feature is indeed the direct consequence of the spherical symmetry of the potential, which allows us to express all kinematical quantities as functions of the single variable $r$ (apart from the various constants) : see Eqs. (\ref{vitesse}) and (\ref{teta}).

\vv \nin The polar equation $r(\theta)$ of a closed orbit is necessarily a periodic function where  $\theta$ should be considered as varying from $-\infty$ to $+\infty$. If the angular period of the motion is larger than $2 \pi$, the curve of the orbit makes more than one tour around $O$ during a period, and inevitably intersects itself in at least one double point at an intermediate distance between $r_m$ and $r_M$. It can be shown that such double points also define symmetry axes of the orbit : see Appendix A. 

\vv \nin Consider a trajectory confined in a circular crown $r_m \leq r \leq r_M$ and find out on which condition it is closed. Because $\dot{\theta} >0$, the particle turns indefinitely around the source $O$ and its trajectory is alternately tangent to the circles of radius $r_m$ and $r_M$ at turning points. When moving from a turning point $P_1$ at $r=r_m$ to the next  turning point $P_2$ at $r=r_M$, the vector ${\bms r}$ rotates by the angle 

\beq \Delta \theta = \di{L \over{\sqrt{2m}} }\, \di{\int^{r_M}_{r_m}} \, \di{{dr}\over{r^2 \sqrt{E-U(r)}}} \label{deltet} \enq

\vv \nin see (\ref{teta}), and rotates by the {\it same} angle when moving from point $P_2$ to the next turning point $P_3$ at $r=r_m$, so that the angle between the two vectors ${\bms {OP}}_1$ and ${\bms {OP}}_3$ is $2 \Delta \theta$.  The trajectory is closed if after an integer $M$ of such rotations the final turning point is the same as the first one. As these rotations are equivalent to a rotation of angle $2 \pi Q$ around $O$, $Q$ being integer, we must have $ 2 M \Delta \theta=  2 \pi Q$ or $\Delta \theta = Q \pi/M$, which means that the trajectory is closed if and only if $\Delta \theta$ is commensurable with $\pi$. It has been demonstrated by Bertrand, Ref. \cite{BERT}, that only two attractive potentials fulfill this requirement {\it at any distance from the source and for any corresponding permissible values of $E$ and $L$}, with  (mod. $\pi$) $\Delta \theta = \pi$ for the Newtonian potential $V(r) = -K_N/r$ ($K_N >0$) and $\Delta \theta =\pi/2$ for the Hookean potential $V(r) = K_H r^2/2$ ($K_H>0$). For these two cases only, all bound trajectories are closed. 

\vv \nin A closed orbit has, at least, one couple of turning points, and the separation angle $\Delta \theta$ between their corresponding axes of symmetry is certainly commensurable with $\pi$. Note that if the tangent is required to be unique in each point of the curve, the period must be exactly $2\pi$. The orbit is then realized in only one tour and is a simple curve. If $c \geq 1$ is the number of couples of turning points like $(P_1,P_2)$ defined above, we then have $c(2 \Delta \theta) = 2 \pi$, and $\Delta \theta = \pi/c$.    

\vv \nin Further, it can be shown that the condition for a potential to provide, at any distance from the source, stable circular orbits and closed quasi-circular trajectories, is to be of the form 
$V(r) = a / r^\eta $, with $ \eta = 2-\gamma^2 < 2$, $\gamma$ being a rational number, and $a \,\eta <0$ : see Ref. \cite{GSP} \S3.6. For these closed quasi-circular orbits, we have $\Delta \theta = \pi/\sqrt{2-\eta}$. Thus, consider first an attractive potential of the form $V(r) = - K/r^\alpha$ with $K >0$ and $\alpha >0$. If $\Delta \theta$ is to be an intrinsic property of this potential, independent of the parameters of orbits, we can compute Eq. (\ref{deltet}) in the limit $E \rightarrow -0$, see Refs. \cite{BERT}, \cite{ARN}. The result is, see Ref. \cite{Car3},   

\beq \Delta \theta = \di{\int^1_0} \di{{dx} \over{\sqrt{x^\alpha - x^2}}} = \di{\pi \over {2- \alpha}}     \label{arn1} \enq

\vv \nin and is valid only if $\alpha <2$. We thus get $1 \leq c = 2-\alpha <2$.  Then $c=1= \alpha$ and $\Delta \theta = \pi$. In the case of potentials such as $V(r) = K^\prime r^\zeta /2$ with $K^\prime >0$, $\zeta >0$, the computation of Eq, (\ref{deltet}) in the limit $E \rightarrow + \infty$ leads to 

\beq \Delta \theta = \di{\int^1_0}\, \di{{dx}\over{\sqrt{1-x^2}}} = \di{\pi \over 2} 
\label{arn2} \enq 

\vv \nin which now gives $c=2= \zeta$. Hence the conclusion of Bertrand's theorem. It is worth noticing here that the values Eq. (\ref{arn1}) and (\ref{arn2}) are obtained exactly from (\ref{deltet}) without any approximation, for the respective potentials $-K/r$ ($\alpha =1$) and $K^\prime r^2 /2$ ($\zeta = 2$), while both values of $\Delta \theta$ are only reached in the  limits defined above for other corresponding potentials. This reinforces the above conclusion.  

\vv \nin Finally, let us mention that a good idea of the shape of trajectories can also be obtained using the well known formula 

$$ \kappa = - \di{L \over{m^2 r v^3}}\, {\cal F}(r),~~{\rm with}~~~{\cal F}(r) = - \di{{dV}\over{dr}}(r)  \label{courbure} $$

\vv \nin giving the curvature $\kappa$ of a trajectory : depending on whether the force is attractive or repulsive everywhere, the curvature is always positive or negative, respectively.

% ==============================================
\section{\large How to characterize a hidden symmetry ?}

\setcounter{equation}{0}
\renewcommand{\theequation}{\mbox{3.}\arabic{equation}}

\vv \nin A dynamical or hidden symmetry must be understood as a continuous group of transformations linking solutions of a set as wide as possible. As shown in Ref. [{\bf 5}], such a symmetry, if it exists, can be characterized in a minimal way by a unitary vector ${\bms U}$, called there the {\it eccentricity} vector, but which we will now call more appropriately a {\it symmetry vector}. Note that the authors of Refs. \cite{BRS,FRD,MUK} used the same formalism as our. Their symmetry vector is currently called the Fradkin-Bacry-Ruegg-Souriau's perihelion vector. A symmetry vector is a first integral that does not depend explicitly on time and can be used to construct with $H$ and ${\bms L}$ a complete set of  independent first integrals of the problem. The role of a symmetry is to provide continuous transformations of a trajectory into another one of the same species by means of Poisson brackets, while defining in some way a common geometrical property of those trajectories, and the vector ${\bms U}$ has to be chosen in this perspective. The motion  being planar, that vector is more likely in the same plane, i.e., orthogonal to ${\bms L}$. With this requirement, we have shown in Ref. [{\bf 5}] that the components of ${\bms U}$ have the simple Poisson brackets, see also Ref. \cite{MUK}  

 \beq \{H, U_i\}= \di{{\partial H }\over{\partial x_r}} \di{{\partial U_i }\over{\partial p_r}}-  \di{{\partial H}\over{\partial p_r}} \di{{\partial U_i }\over{\partial x_r}} = 0,~~\{L_i, U_j \} = \ep_{ijk} U_k,~~\{ U_i, U_j\} =0 \label{PB1} \enq 

\vv \nin the indices $i, j, k$ taking each the values $1,2,3$ standing for $x,y,z$, respectively ; $\ep_{ijk}$ is the complete skew-symmetric Levi-Civita tensor for three dimensions, with $\ep_{123}=1$. Choosing a symmetry vector as unitary has the advantage to account for any kind of dynamical symmetry, and so could be considered as the true indicator of a symmetry of trajectories. As an operator in phase space ${\bms U}$ is to be considered as a field vector in that space, and treated in exactly the same way as the angular momentum ${\bms L}$. On the other hand, at a given point in phase space, the geometrical role of ${\bms U}$ is to define a symmetry axis of the trajectory corresponding to that point.   

\vv \nin Next, let us define ${\bms W} = \di{1\over L} {\bms L} \times {\bms U}$ which is a vector forming with ${\bms U}$ and ${\bms \ell} = {\bms L}/L$ an orthonormal basis. It is well known that  the components of the vectors of such a basis are linked by the formulas   

$$ W_i = \ep_{ijk} \ell_j U_k,~~ U_i = \ep_ {ijk} W_j \ell_k,~~\ell_i = \ep_{ijk} U_j W_k, $$
\beq  \ep_{ijk} U_k = W_i \ell_j - W_j \ell_i,~~ \ep_{ijk} W_k = \ell_i U_j - \ell_j U_i,~~\ep_{ijk} \ell_k = U_i W_j - U_j W_i,    \label{ORT} \enq
$$ U_i U_j + W_i W_j + \ell_i \ell_j = \delta_{ij}  $$

\vv \nin The vector ${\bms W}$ being also orthogonal to ${\bms L}$, its components satisfy 

$$ \{L_i, W_j \} = \ep_{ijk} W_k = \ell_i U_j - \ell_j U_i ,~~\{W_i, W_j\} =0$$

\vv \nin To express the transformation of $L$ under the action of ${\bms U}$, we need to compute some Poisson brackets. From Eq. (\ref{PB1}),  we get   

\beq  \{L, U_j\} = \di{1\over {2L}} \{ L^2, U_j \}= \di{1\over{2 L}} \{ L_i L_i, U_j\} = \di{ L_i \over L} \{L_i, U_j \}  = \ep_{ijk} \ell_i U_k = - W_j  \label{PB2} \enq

\vv \nin In the same way, we obtain

$$  \{L, W_j\} = \di{ L_i \over L} \{L_i, W_j \}  = \ep_{ijk} \ell_i W_k =  \ep_{jki} W_k \ell_i =U_j  \label{PB3} $$

\vv \nin Using Eqs. (\ref{PB2}) and (\ref{ORT}),  we have 

$$ \{ \ell_i, U_j \} = \{ L_i, U_j\}/L - L_i \{L, U_j\}/L^2 = \ep_{ijk}U_k/L + L_i W_j/L^2 =W_i L_j/L^2 $$

\vv \nin whence 

$$ \{ U_i, W_j \} = \ep_{j mn} \{U_i , \ell_m U_n\} = \ep_{jmn} U_n \{ U_i, \ell_m \} $$
$$ = -\ep_{jmn} U_n W_m L_i/L^2 = L_i L_j/L^3 $$

\vv \nin Under an infinitesimal transformation generated by a symmetry vector ${\bms A}({\bms r}, {\bms p})$, the infinitesimal variation of $L$ is given by 

\beq \delta L = \{ \alpha_i A_i, L\} \label{VI}\enq

\vv \nin where $\alpha_i$ are the components of an infinitesimal constant vector ${\bms \alpha}$ (i.e., independent of the canonical variables). From the dimensional analysis of Eq. (\ref{VI}), $\alpha_i A_i$ must be homogeneous to an angular momentum. If ${\bms A} = {\bms U}$, then the $\alpha_i$ are to be homogeneous to $L$. If the $\alpha_i$ are chosen dimensionless, ${\bms A}$ must be homogeneous to $L$ and for this reason cannot be a unitary vector. At this stage, $L$ is the only first integral that is homogeneous to an angular momentum, and we are naturally led to define a new symmetry vector by ${\bms A} = L\, {\bms U}$. With this choice, we obtain the new set of relations 

\beq \{L_i, A_j \}= \ep_{ilk} A_k,~~\{A_i, A_j\}= - \ep_{ijk} L_k,~~\{L_i, L_j\} = \ep_{ijk} L_k \label{lieal} \enq

\vv \nin which is homomorphic to that defining the Lie algebras of the groups $SO(3,1)$, or the Lorentz group $L(3,1)$ or the group $SL(2,C)$ : see, e.g., Ref. \cite{BAC}. Indeed, it appears that the latter choice yields a simpler law of transformation for $L$, as seen below. Defining ${\bms B} = L \,{\bms W}$,  we have 

\beq \{L_i, B_j\} = \ep_{ijk} B_k,~\{B_i, B_j\}= -\ep_{ijk} L_k,~   
 \{ A_i, B_j\} = L\, \delta_{ij} \label{AB1} \enq 

\vv \nin With finite $\alpha_i$, the transformation ${\cal T}_A  $ generated by $T_A = {\bms \alpha}\cdot{\bms A}$ changes $L$ into 

$$ L^\prime = {\cal T}_A  (L) = L + \{T_A, L \} + \di{1\over 2} \{T_A, \{T_A,L\} \} + \di{1\over 3!} \{T_A, \{T_A, \{T_A,L\}\}\} +\cdots $$

\vv \nin Let us define $\xi = ||{\bms \alpha}||$, ${\bms n} ={\bms \alpha}/\xi$. Then, 

$$ T_A(L) = \{T_A, L\}= \alpha_i \{A_i, L\} =  {\bms \alpha} \cdot {\bms B} = \xi \,L\, {\bms n}\cdot {\bms W}, $$
$$ T^2_A(L) = \{T_A, \{T_A, L\}\} = \alpha_i \alpha_j \{A_i, B_j\} = \xi^2 L, $$
$$ T^3_A(L) = \{T_A,\{T_A,\{T_A,L\}\}\}= \xi^2 \{T_A,L\} = L\, \xi^3 {\bms n}\cdot {\bms W} $$

\vv \nin and so on. Hence, we finally obtain the law of transformation for $L$

\beq L^\prime = L \left[ \cosh \xi + {\bms n}\cdot{\bms W} \sinh \xi  \right] \label{trL} \enq 

\vv \nin which is a simple scaling. Let us remark that, generally speaking, the only invariants under the transformations generated by the Lie algebra Eq. (\ref{lieal}) are the two Casimir operators $C_1 = L^2 - A^2$ ($A^2 = A_i A_i$) and $C_2 = {\bms L}\cdot {\bms A}$, which are both zero here. Thus, $L^2$ and $A^2$ are not invariants, only their difference is. This explains the result $L^{\prime} \neq L$ in Eq. (\ref{trL}).      

\vv \nin Note also that any operator like $T_A$ acts as a differentiation (it verifies Leibniz rule). Consequently, $F_1$, $F_2$, $F_3\circ F_1$ being any regular functions of canonical coordinates  its exponential ${\cal T}_A  $ has the two important properties :    

$$ {\cal T}_A  (F_1 F_2) = {\cal T}_A  (F_1) {\cal T}_A  (F_2),~~{\rm and}~~ {\cal T}_A  ( F_3(F_1) ) = F_3\left( {\cal T}_A  (F_1) \right)  $$ 

\vv \nin which in particular ensure the covariance under ${\cal T}_A  $ of the common polar equation of trajectories. For example, in the Kepler problem, they allow us to write that the eccentricity of an ellipse with parameters $E$ and $L$ is transformed into the eccentricity of another ellipse with parameters $E$ and $L^\prime = {\cal T}_A  (L)$. 

\vv \nin Let us also give the transformation of the components of ${\bms L}$ and ${\bms A}$. We have 

$$ T_A(L_i) = \alpha_j \{A_j, L_i\}  = \xi n_j \ep_{jik} A_k  = \xi \left[ L_i \,{\bms n}\cdot {\bms W} - W_i \,{\bms n}\cdot{\bms L} \right], ~~{\rm whence}  $$
$$  \{ T_A, {\bms n} \cdot {\bms L} \} =0 $$
$$  T^2_A(L_i) = \xi^2  n_\ell n_j \ep_{jik}\{A_\ell, A_k \} = \xi^2 n_j n_\ell \ep_{jik} \ep_{\ell m k} L_m = \xi^2 \left[ L_i - n_i \, {\bms n} \cdot {\bms L}     \right]   $$

\vv \nin where the formula $\ep_{jik}\ep_{\ell m k} = \delta_{j \ell} \delta_{im} - \delta_{jm} \delta_{i \ell}$ has been used in the last equation. Then, 

$$ T^3_A(L_i) = \xi^2\, T_A(L_i), ~~T^{2n+1}_A(L_i) = \xi^{2n} \,T_A(L_i)~~{\rm with}~~ n \geq 1, $$
$$T^4_A (L_i) = \xi^2\, T^2_A(L_i),~~T^{2n} (L_i) = \xi^{2n-2}\, T^2_A(L_i)~~{\rm with}~~n \geq 2, $$

\vv \nin and finally,  

$$ L^\prime_i = {\cal T}_A  (L_i) = L_i + \left( \cosh \xi -1 \right)  \left[ L_i - n_i \, {\bms n} \cdot {\bms L}     \right]  + \sinh \xi \, \left[ L_i \,{\bms n}\cdot {\bms W} - W_i \,{\bms n}\cdot{\bms L} \right] $$
\beq  = L_i \left[ \cosh \xi + {\bms n}\cdot {\bms W} \, \sinh \xi  \right] - {\bms n} \cdot {\bms L} \left[ n_i (\cosh \xi -1) + W_i \sinh \xi \right] \label{trli} \enq

\vv \nin The transformation of the components $A_i$ is obtained from Eq. (\ref{trli}), simply replacing $L_i$ and ${\bms L}$ with $A_i$ and ${\bms A}$. Then, using ${\cal T}_A  (A_i) = {\cal T}_A  (L U_i) = {\cal T}_A  (L) {\cal T}_A  (U_i)$, we find 

$$  A^\prime_i = A_i \left[ \cosh \xi + {\bms n}\cdot {\bms W} \, \sinh \xi  \right] - {\bms n} \cdot {\bms A} \left[ n_i (\cosh \xi -1) + W_i \sinh \xi \right] ~~~{\rm and}  $$
\beq U^\prime_i = U_i -{\bms n} \cdot {\bms U} \, \di{{ n_i \left( \cosh \xi -1 \right)   + W_i \sinh \xi }\over{  \cosh \xi + {\bms n}\cdot{\bms W} \sinh \xi  }} \label{trui} \enq

\vv \nin In the same way, it is easy to obtain 

$$ B^\prime_i = B_i + n_i \left(\cosh \xi -1 \right) \, {\bms n}\cdot {\bms B} + n_i L \sinh \xi ~~~{\rm and} $$
\beq W^\prime_i = \di{{ W_i + n_i \left( \cosh \xi -1 \right)  \, {\bms n} \cdot {\bms W}   + n_i \,\sinh \xi }\over{  \cosh \xi + {\bms n}\cdot{\bms W} \sinh \xi  }} \label{trwi} \enq

\vv \nin taking into account the relations $\{ A_i, B_j\} = L \delta_{ij}$ (Eq. (\ref{AB1})) and ${\cal T}_A  (B_j)= L^\prime {\cal T}_A  (W_j)$. 

\vv \nin As shown in Appendix B, Eq. (\ref{trwi}) can be interpreted as a Lorentz boost. However, the orthonormality of the reference basis ${\bms U}, {\bms W}, {\bms \ell}$ is preserved under ${\cal T}_A$, Eqs. (\ref{trli}), (\ref{trui}) and (\ref{trwi}) defining the new reference basis ${\bms U^\prime}, {\bms W^\prime}, {\bms \ell^\prime}$. Hence, ${\cal T}_A$ generally changes the  orientation of the plane of motion, a change that should be described by a rotation whose axis and angle are to be expressed in terms of ${\bms n}$ and $\xi$. This is also done in Appendix B. In addition, it is seen from Eqs. (\ref{trL}),  (\ref{trli}) and (\ref{trui}) that ${\cal T}_A  $ also causes a change of scale along the reference axes defined by ${\bms A}$, ${\bms B}$ and ${\bms L}$. As they have the same norm $L$, the latter are multiplied by the same factor $\rho = L^\prime/L= \cosh \xi + {\bms n}\cdot{\bms W}\, \sinh \xi$. Thus, the action of ${\cal T}_A$ on the said three vectors is equivalent to the product of a rotation ${\cal R}_A$ by $\rho$. If ${\bms n}$ is along the initial ${\bms W}$, i.e., if ${\bms n} = \pm {\bms W}$, the reference basis remains unchanged while the scale factor is $\exp(\pm \xi)$.

\vv \nin Unfortunately, we cannot derive a general form of the transformation of the coordinates $X$ and $Y$ as such a transformation is specific to the dynamics of the motion. Actually, the previous developments are all we can do independently of the particular  
(spherical symmetric) potential under study. 

\vv \nin As emphasized in Ref. [{\bf 1}], the transformations of dynamical symmetries alone are unable to link trajectories corresponding to different values of energy, because their generators  have zero Poisson brackets with the Hamiltonian. As shown in the same reference, this lack is overcome for potentials in single power law, thanks to their additional symmetry called ``mechanical similarity". This symmetry allows us to complete the group of transformations that is necessary to obtain a full connection between trajectories of a given set by varying both variables $E$ and $L$. Finding an equivalent symmetry for other kinds of spherically symmetric potentials is an open question.

% ==============================================
\section{\large The choice of reference axes}

\setcounter{equation}{0}
\renewcommand{\theequation}{\mbox{4.}\arabic{equation}}

\vv \nin Thanks to rotational invariance, the reference axes in the plane of motion can be freely chosen. However, on one hand, we all know that whatever the problem, the best way to present its solution in the most intelligible and transparent form is to take into account from the outset  some of their common features, if any. On the other hand, we are convinced that a dynamical symmetry, as currently expressed mathematically, is a direct consequence of the existence of common geometrical characteristics of a set of trajectories. Hence, we are immediately led to look for such characteristics and, of course, accounting for the previous section, a possible axis of symmetry that could serve as a reference. Before that, it is necessary to point out some subtleties. 

\vv \nin First, we must mention a fact that seems obvious but is essential in our study : {\it whatever the shape} of the potential $V(r)$ and the range of variations of $r$, it is always possible to find a value of $E$ satisfying the inequality $E \geq U(r)$ (defining the allowed motions), with the consequence that the equation $E=U(r)$ of turning points has at least one solution for $r$ which, unfortunately, is hard or impossible to obtain in closed form in the general case. Using elementary arguments, the existence of turning points can be proved for the attractive power-law potentials $-K/r^\alpha$ and $K r^\alpha/2$ ($K >0$, $\alpha >0$), and for power-law potentials that are everywhere repulsive. As explained in Appendix A, every turning point is associated with an axis of symmetry, thanks to spherical symmetry. Thus, the search for symmetry axes requires that of turning points. This can be done through
 the polar equation $r(\theta)$ one derives from Eqs. (\ref{eqr}) or (\ref{teta}), the parameters of which are determined by the values of $E$ and $L$. Giving to $\theta$ all permissible values, this polar equation represents the complete geometrical curve along which take place all motions having initial conditions that are different but give the same values for $E$ and $L$. This curve  will be also called a {\it trajectory} hereafter. The said motions can be linked by varying continuously their initial conditions, hence the said polar equation already covers an infinity of possible motions. Due to this diversity of initial conditions, it may happen that some of these motions do not have a turning point, and no symmetry axis, whereas, according to the above  remark, such a point does exist and may be present on the complete geometrical curve (it corresponds to $dr/d\theta =0$). It may also happen that, depending on the values of $E$ and $L$, some trajectories do not have a symmetry axis. The following positive arguments can be set against this last negative remark. As we know, the existence of an axis of symmetry for a class of trajectories implies that of additional first integrals. But the latter, when expressed in terms of canonical variables, owe their status solely to the fact that their Poisson brackets with the Hamiltonian are zero : they are not attached to the class of trajectories from which they have been highlighted, they apply to all possible trajectories and motions. In addition, for a given potential, all  possible trajectories are deduced from the same fundamental equation and must have similar or analytically related polar equations, and these necessarily bear a trace of the existence of a turning point. 

\vv\nin Then, consider a trajectory (complete geometrical curve) having a turning point $P_0$, where the particle has 
the vector position ${\bms r}_0 = r_0 \,{\bms e}_{r0}$ and momentum ${\bms p}_0 = p_0 \,{\bms e}_{\theta 0}$ such that ${\bms r}_0 \cdot {\bms p}_0 =0$. Its axis of symmetry is along  ${\bms r}_0$ and passes through $O$. In the 2-plane of motion, we redefine the unit vectors of rectangular axes as 
${\bms E}_x = {\bms e}_{r0}$ and ${\bms E}_y = {\bms e}_{\theta0}$. Note that the angular momentum is completely defined by ${\bms r}_0$ and ${\bms p}_0$, as $L= r_0 p_0$. Of course, the two unit vectors ${\bms E}_x$ and ${\bms E}_y$ are constant in time and, as such, are good candidates to represent a (non-zero) symmetry vector. Let us choose ${\bms U} = {\bms E}_x$. To make this vector a field vector in phase space, we have to express it in terms of the  canonical variables ${\bms r}$ and ${\bms p}$ of a point of the trajectory, in the same way the angular momentum is written as ${\bms L} = {\bms r} \times {\bms p}$. This is easily done by simply inverting the trivial expressions 

$$ {\bms r} = X {\bms E}_x + Y {\bms E}_y ,~~~{\bms p} = P_x {\bms E}_x + P_y {\bms E}_y \label{triv1} $$  

\vv \nin Using 

$$ L= X P_y - Y P_x,~~P_y = p_r \sin \psi + p_\psi \cos \psi = \di{u \over r} \sin \psi + \di{L\over r} \cos \psi \label{triv2} $$ 

\vv \nin where $\psi$ is the polar angle of ${\bms r}$ relatively to ${\bms r}_0$, as defined in Eq. (\ref{psi}), we get 

$$ {\bms E}_x = \di{1\over L} \left[ P_y \,{\bms r} -Y \,{\bms p} \right],~~{\rm or} $$
\beq {\bms E}_x = \di{1\over{ L r}} \left[ \left( u \sin \psi + L \cos \psi \right) {\bms r} - r^2 \sin \psi\, {\bms p}   \right]  \label{ux} \enq   

\vv \nin The second member of Eq. (\ref{ux}) is a function of canonical variables, $\psi$ being expressed as in Eq. (\ref{psi}) with $L = \sqrt{p^2 r^2 -u^2}$ and $E = p^2/(2m) + V(r)$. It is obviously a first integral as ${\bms E}_x$ is a constant of motion. Note that $X$, $Y$, $P_x$,  $P_y$ and $\psi$ are to be considered now as invariant under rotations, expressed in terms of the scalars  $r^2$, $p^2$ and $u = {\bms p}\cdot{\bms r}$. Using 

$$ {\bms p} = \di{1\over r^2} \left[\, u {\bms r} - {\bms r} \times {\bms L}\,\right] $$ 

\vv \nin and the polar equation $r = R(\psi)$, Eq. (\ref{ux}) can be recast in the useful form 

\beq {\bms E}_x = \di{1\over{ L R(\psi)}} \left[  \sin \psi \, {\bms r} \times {\bms L} + L \cos \psi  \,{\bms r}   \right]  \label{ux2} \enq   

\vv \nin Instead of obtaining a trivial relation when projecting ${\bms r}$ onto ${\bms E_x}$, this trick allows us to ``retrieve" the polar equation : 

$$ r \cos \psi = {\bm r} \cdot {\bms E_x} = \di{{r^2 \cos \psi}\over R(\psi)},~~{\rm hence}~~r = R(\psi) $$

\vv \nin At this point, it is worth remarking that the expressions in Eqs. (\ref{ux}) and (\ref{ux2}) are covariant under the symmetry group, a property that could not have been obtained using as a reference any other constant vector independent of the dynamics. 

\vv \nin Defining an axis of symmetry, ${\bms E_x}$ must play a role similar to that of the LRL vector in the Kepler problem ($V(r) = -K/r$). Let us check Eq. (\ref{ux}) by considering a few   examples. The first one is just that of the Kepler problem. In the case of ellipses, we have 

$$ R(\psi) = \di{r_0 \over{1 + e \cos \psi}} , ~{\rm with}~~r_0 = \di{L^2\over{mK}},~e = \sqrt{1 - 2 r_0|E|/K},~u = e L r \sin \psi/r_0  $$ 

\vv \nin and we find 

\beq  K e \, {\bms E}_x = - \di{u\over m}\, {\bms p} +\left(\di{p^2\over m} - \di{K\over r} \right) {\bms r} \label{VLRL} \enq  

\vv \nin which is exactly the LRL vector. The same result is obtained considering instead a hyperbola. 

\vv \nin The second example is that of the Hookean (or harmonic) potential $V(r) = K r^2/2$ with $K>0$. The more general trajectory is an ellipse centered in $O$, whose cartesian equation 
is $X^2/a^2 + Y^2/b^2 =1$. It appears more convenient to use the parametrisation with the eccentric anomaly $\chi$ : $X = a \cos \chi,~Y= b \sin \chi $, which leads to 

$$ \dot{\chi} = \sqrt{\di{K \over m}},~~L = ab \sqrt{Km},~~P_y = b \sqrt{Km}\, \cos \chi, $$
\nin and 
$$ {\bms E}_x = \di{1\over{a \sqrt{Km}}} \left[ \sqrt{Km} \cos \chi\,{\bms r} - \sin \chi \,{\bms p} \right]    \label{hux}  $$ 

\vv \nin Using the notations of Ref. [{\bf 5}], we find 

$$ \cos \theta = \di{{p^2 - Km r^2}\over{\sqrt{(p^2 - Km r^2)^2 + 4 Km u^2}}} = \cos 2 \chi , 
$$
$$ \sin \theta = \di{{2u \sqrt{Km}}\over{\sqrt{(p^2 - Km r^2)^2 + 4 Km u^2}}} = \sin 2 \chi $$

\vv \nin Then, by setting $\lambda = Km a^2$, $\chi = \theta/2$, we retrieve the opposite of one of the two ``eccentricity vectors" proposed in Ref. [{\bf 5}] for this case. 

\vv \nin These two examples demonstrate the efficiency of this simple method in finding the expression of symmetry vectors in terms of canonical variables, once the polar equation of trajectories is known. 

\vv \nin Let us now consider the instructive example of the trajectories provided by the potential $V(r) = -K/r^2$. Assuming $L^2 \neq 2mK$, their polar equations can be derived from the general expression   

$$  R(\theta) = \di{ R \over \cos \chi (\theta - \phi)}~~~{\rm with}~~\chi^2 = 1 - \di{{2m K}\over L^2},~~ R^2 = L \chi^2 /(2mE) \label{r2pe} $$

\vv \nin $\theta$ being the polar angle of ${\bms r}$ relatively to its initial value ${\bms r}(0)$ and $\phi$ being constant. Note that $\chi$, $R$ and $\phi$ are real or complex numbers according to the signs of $\chi^2$ and $E$, subject to the condition $R(\theta) >0$. There are three main cases : 

\vv \nin 1) $\chi^2 >0$, $E>0$ ; then 

$$ R(\theta) = \di{r_0 \over \cos \omega (\theta - \phi)}~~~{\rm with}~~\omega = \sqrt{1 - \di{{2mK}\over L^2}}, ~~r_0 = \di{{L \omega} \over{\sqrt{2mE}}},~~\phi ~{\rm real}  $$  

\vv \nin 2) $\chi^2 <0$, $E<0$ ; then 

$$ R(\theta) = \di{r_0 \over \cosh \Omega (\theta - \phi)}~~~{\rm with}~~\Omega = \sqrt{ \di{{2mK}\over L^2} -1}, ~~r_0 = \di{{L \Omega} \over{\sqrt{2m|E|}}},~~\phi ~{\rm real}  $$  

\vv \nin 3) $\chi^2 <0$, $E>0$. Then $\chi =\pm i \Omega$, $R= \pm i r_0$ with $r_0 = \di{{L \Omega} \over{\sqrt{2m E}}}$, and we are led to set $ \Omega \phi = \pm \Omega \phi^\prime - i \pi/2$ with $\phi^\prime$ real and positive, to take into account all possible initial conditions. This gives  

\beq R(\theta) = \di{{\pm r_0} \over \sinh \Omega (\theta \pm \phi^\prime)} \label{cas3} \enq  

\vv \nin Case 1) is mainly that of unbounded trajectories. The complete geometrical curve is obtained by varying $\psi= \theta - \phi$ between $- \pi/(2 \omega)$ and $+ \pi/(2 \omega)$. It has a single turning point at $\psi =0$, that is reached or not during a real motion according to whether the initial angle $\theta(0)$ is less or greater than $\phi$ (remember that $\dot{\theta} >0$). Independently of this possibility, the polar equation is symmetrical relatively to the value $\theta = \phi$, corresponding to an axis of symmetry which is opportunately taken as axis $Ox$. In this case, we find the symmetry vector 

$$ {\bms E}_x = \di{{\cos \omega \psi}\over{ L r_0}} \left[  \sin \psi \, {\bms r} \times {\bms L} + L \cos \psi  \,{\bms r}   \right],~~ {\rm with}~~ \psi = \pm \di{1\over \omega} \cos^{-1} \left(\di{r_0 \over r}\right) \label{alf2ub} $$

\vv \nin  Case 2) mainly corresponds to bounded trajectories. In a real motion, the particle always falls towards the source in a spiral movement. The complete geometrical curve is obtained by varying $\psi = \theta - \phi$ between $-\infty$ and $+ \infty$. It has also a single turning point at $\psi=0$.  Even for this special geometry, we have a symmetry vector : 

$$ {\bms E}_x =\di{{\cosh \Omega \psi}\over{ L r_0}} \left[  \sin \psi \, {\bms r} \times {\bms L} + L \cos \psi  \,{\bms r}   \right] ~~ {\rm with}~~ \psi = \pm \di{1\over \Omega}\cosh^{-1} \left(\di{r_0 \over r}\right) \label{alf2b} 
$$

\vv \nin Case 3) gathers two subcases. The first one with the plus sign in Eq. (\ref{cas3}) corresponds to bounded trajectories falling spirally towards the source as $\theta$ varies from 0 to $+\infty$, according to the polar equation $r = r_0/\sinh \Omega( \theta + \phi^\prime)$. The second one with the minus sign in Eq. (\ref{cas3}) is that of unbounded trajectories, $r$ going from $r(0)$ to $+ \infty$ as $\theta$ varies from 0 to $\phi^\prime$, according to the polar equation $r = r_0/\sinh \Omega (\phi^\prime - \theta)$. None of these trajectories have a turning point, nor their complete geometrical curves for which $\psi = \phi^\prime + \theta$ for the first one varies between 0 and $+\infty$ and $\psi = \theta - \phi^\prime$ for the second one varies between $-\infty$ and 0. In each case, the value $\psi=0$ refers to an asymptote which could be considered as a reminiscence of the axis of symmetry observed for case 2) when transforming a $\cosh$ into a $\sinh$ in the polar equation, as seen above.  These asymptotes are now taken as reference axes and we have 

$$  {\bms E}_x =\di{1\over{ L r}} \left[  \sin \psi \, {\bms r} \times {\bms L} + L \cos \psi  \,{\bms r}   \right]  ~~~{\rm with}~~~\psi = \pm \di{1\over \Omega} \sinh^{-1} \left(\di{r_0 \over r}\right) $$

\vv \nin This example shows that, as a function of canonical variables, a symmetry vector may have different analytical forms according to the trajectories to which it refers. This fact could explain those singularities of the symmetry vector which have been predicted by various authors, see e.g. Ref. \cite{LEFL,GBM}. 

\vv \nin Another instructive example is provided by the potential 

$$ V(r) = - \di{K_1 \over r} + \di{K_2 \over r^2},~~K_1 >0,~K_2 >0,  \label{reed1} $$ 

\vv \nin which has been studied in a particular context in Ref. \cite{REED}. The corresponding effective potential  

$$ U(r) = \di{L^{\prime 2} \over{2m r^2}} - \di{K_1 \over r},~~{\rm where}~~L^{\prime 2} = L^2 + 2 m K_2  \label{reed2} $$ 

\vv \nin  differs from that of the Kepler problem only by the replacement of $L$ by $L^\prime$. However, we still have $\dot{\psi} = L/(mr^2)$. Consequently, the polar equation of trajectories is found to be 

\beq  R(\psi) = \di{ r_0\over{1 + e \cos (\beta \psi) }}, \label{reed3} \enq 
\nin with $ \beta = L^\prime/L >1,~~r_0 = L^{\prime 2}/(mK_1),~~e = \sqrt{1 + 2E L^{\prime 2}/(mK^2_1)} $.

\vv \nin Bounded trajectories exist for $E<0$. They are closed only for discrete values of $L$ for which $\beta$ is a rational number larger than 1. Otherwise, they are not closed and, as time passes, the associated motion fills  the entire circular crown $r_0/(1+e) \leq r \leq r_0/(1-e)$. Depending on the case, we encounter a finite or infinite number of turning points on one 
tour. Obviously, we can choose any of those on the circle $r = r_0/(1+e)$ (pericenters) to define the reference axis for which we take  $\psi=0$ and obtain the polar equation in the form (\ref{reed3}) with $- \infty < \psi < + \infty$. For $E>0$, we have unbounded trajectories. The only property shared between all these curves is the symmetry of their polar equation, when $\psi$ is replaced by $- \psi$, which means that the axis $Ox$ is their common axis of symmetry. Using Eq. (\ref{ux2}), we find for this problem the symmetry vector 

$$ {\bms E}_x = \di{{1+ e \cos \beta \psi}\over{ L r_0}} \left[  \sin \psi \, {\bms r} \times {\bms L} + L \cos \psi  \,{\bms r}   \right] \label{reed4} $$

\vv \nin Unbounded trajectories have the same geometry, and the dynamical symmetry operations can transform continuously any of them into another having the same value of energy. 

\vv \nin Let us now consider central potentials which are {\it everywhere repulsive}. All corresponding trajectories are unbounded but bear each a sole turning point associated with an axis of symmetry. Even this case has a symmetry group as illustrated by the following two examples. 

\vv \nin The repulsive newtonian potential $V(r) = K/r$ with $K >0$ yields hyperbolas of polar equation  
$$ r= R(\psi) = \frac{r_0}{ e \cos \psi -1} $$

\nin with $r_0 = L^2/(mK)$, $e = \sqrt{1 + 2 E L^2/(mK)}$. This case has the LRL-like vector 

$$
Ke{\bms E}_x=-\frac um{\bms p}+\Big(\frac{p^2}m+\frac Kr\Big){\bms r}
$$

\nin obtained from Eq.~\ref{VLRL} by simply replacing $K$ with $-K$. 

\vv \nin The ``repulsive harmonic" potential $V(r) = - K r^2/2$ with $K>0$ also yields hyperbolas of polar equation 

 $$ R(\psi) = \frac{r_0}{\sqrt{1 +  e \cos 2\psi} }$$

\nin with $r_0 = L/\sqrt{mE}$, $e= \sqrt{1+ L^2K/(mE^2)}$. In the (attractive) Hookean case, the symmetry vector is an eigenvector of the constant symmetric two-rank tensor $T_{ij} = m K x_i x_j + p_i p_j$. Here we also find a constant symmetric two-rank tensor : $T_{ij} = - mK x_i x_j +p_i p_j$, one of its two eigenvectors being along the axis of symmetry of hyperbolas. This vector is used to define a symmetry vector, analogous to the eccentricity vector given in Ref. \cite{car2}.

% ==============================================
\section{\large Conclusion}

\vv \nin The main idea that emerges from our study is that the dynamical or hidden symmetries  discovered for the first time for the Newtonian and the Hookean potentials, are not a specific property of the latter, but instead are inherent to all the spherical symmetric potentials. This feature, that seems to have been widely accepted without solid basis after the works of Refs. \cite{BRS,FRD}, must now be consider as well established. In the case of the Newtonian and the Hookean potentials, their true origin seems to have been masked by the fact that those potentials are the only ones providing closed orbits at any distance from the source. Indeed, this important property has often been considered, without any certainty, as the cause of the occurrence of the observed internal symmetries, or vice versa : see Ref. \cite{BAC} \S6.11. Thus, we must dissociate the existence of a dynamical symmetry from that of closed trajectories. This assertion is reinforced by the fact that even central potentials that are everywhere repulsive also provide a dynamical symmetry though they cannot produce any closed trajectory.

\vv \nin Let us review the different arguments leading to our conclusion. 

\vv \nin An immediate consequence of the spherical symmetry of the potential is that, in the 2-plane of motion,  the distance $r$ from the source is finally the single variable that determines two important geometrical characteristics of a trajectory, namely its limits and the existence of turning points, expected for all such potentials. Thanks again to the spherical symmetry, turning points define symmetry axes that may be considered as the true origin of a dynamical symmetry. Let us remark that a symmetry relative to an axis is generally reflected in the expression of the polar equation representing a given set of trajectories, and is therefore a common property of the latter. A natural way to characterize such axis is by means of its unit vector, which has been called here a {\it symmetry vector}. This vector is lying in the 2-plane of motion and is constant in time. For a $n$-dimensional configuration space, its $n-1$ independent components can be chosen as the first integrals forming, together with angular momentum and energy, a complete set of $2n-1$ fundamental and independent first integrals for the motion under study. Poisson brackets between those first integrals yield first integrals that cannot be independent of them. So, we may expect that performing Poisson brackets between fundamental first integrals leads to relations similar to that defining a Lie algebra. Here, the underlying idea is that there must exist a continuous group of transformations, the dynamical group, connecting all trajectories having a common symmetry, and which is generated by the said algebra. This is indeed realized, and the Lie relations obtained with a unit symmetry vector are very simple and apply to any situation. With a slighty modified symmetry vector, the obtained Lie algebra is homomorphic to that of the $SO(3,1)$ group. 

\vv \nin It is worth mentionning here that the axial symmetry resulting from the inoperative change of $\psi$ into $- \psi$ cannot of course be deduced from the equations of motion using the formalism of Lie's transformations, since the latter applies only to continuous symmetries. The dynamical group does not describe directly this operation either, but takes into account the presence of an axis of symmetry for each trajectory. 

\vv \nin To summarize, for the problem under study, what is often called ``hidden symmetry" and expected to be accounted for by a dynamical symmetry group is just a mathematical expression of the symmetry of the trajectories (or at least some of them), which goes beyond the simple case of the Newtonian and the Hookean potentials. Let us also emphsize here the status of a symmetry group as a stabilizer of the common structure of the trajectories under study, Ref. \cite{cari1}.    

\vv \nin Examples have been given, showing that dynamical symmetries are not only present for potential in a simple power law in $r$, just as in the Kepler and the Hookean problems, but also occur in problems with composite (or inhomogeneous) potentials. At this point, It must be remembered that energy, which is one of the parameters defining trajectories, cannot be modified by a dynamical group as defined above. Therefore, one has to enlarge this group in finding, for each case, an additional symmetry providing, if it exists, a complete mapping of trajectories, just as ``mechanical similarity" does for potentials in single power law.   

\vv\nin Let us make some final observations. As is well known, having three independent and Poisson-involutive first integrals, mamely the energy $E$, the norm $L$ of the angular momentum and its component $L_z$, the motion of a particle in a spherical symmetric potential is said to be an integrable problem, in the Liouville sense. But, as shown in this article, for any potential $V(r)$, its spherical symmetry provides us with two additional and independent first integrals. For this reason, the said problem should be re-qualified as maximally superintegrable.

\vv \nin Finally, it should be noted that finding a continuous dynamical group for the system considered in this article appears rather natural, as explained below. An obvious but essential condition for the existence of such a group is the possibility of a continuous (or analytic) link between trajectories (complete geometrical curves). This possibility does exist for our system, and the said link is supposed to be realized mathematically by the transformations of the group, just as canonical transformations. Changing a trajectory into another of the same species remains to change continuously the values of the above-mentioned fundamental first integrals. But the latter being Poisson-involutive, the transformations they generate through Poisson brackets are unable to achieve this change. The rotational group also cannot change the values of $E$ and $L$. The only possible conclusion is that there must exist at least one additional first integral which has non-zero Poisson brackets with the fundamental involutive first integrals and thus generates transformations changing in particular the value of $L$ (changing the value of $E$ is more complicated). In the present case, we have found two new first integrals expanding the action of the rotational group alone while carrying a common geometrical property of trajectories. From this observation, it is also natural to ask whether a dynamical group could exist for any integrable system whose solutions can be linked continuously. If so, such an integrable system would  also be {\it de facto} superintegrable. Then, the simple condition of a continuous link between solutions would appear independent of and, conceptually, more important than the very nature of the system under study, that nature manifesting only in the representation of the group and the degree of superintegrabllity, specific to that system. This last conjecture is also based on the fact that the dynamical group considered in this article can be described by a single formalism, whatever the potential.

\vv \nin All the above observations might be of interest to researchers working on maximally superintegrability, Refs. \cite{TTW,MPW,AMP,MSW,ABG}, and its extension to systems having non-spherical symmetric potentials, see Refs. \cite{POST,BIZ}.

\bibliographystyle{amsplain}

\renewcommand{\refname}{\large{References}}

\newpage

% ==============================================

\nin {\large \bf Appendix A : Axes of symmetry of a trajectory}

\setcounter{equation}{0}
\renewcommand{\theequation}{\mbox{A.}\arabic{equation}}

\vvv \nin In the demonstration of his theorem,  Bertrand started by saying that turning points are defining symmetry axes of trajectories, a statement which, in our opinion, deserves more explanation than that is usually given in textbooks, e.g. in Ref. \cite{LAND} \S14. 

\vv \nin Let us consider a trajectory inside the circular crown $r_m \leq r \leq r_M$, making more than one tour around $O$ and intersecting itself in a double point $P_0$ at the intermediate distance $r_0$. Let $B_a$ and $B_b$ be the two branches of the curve crossing at $P_0$. The polar angle of a radius vector ${\bms r} = r {\bms e}_r$ relatively to ${\bms {OP}}_0 = r_0\, {\bms e}_{r0}$ is, mod. $2\pi$,  

\beq \psi = \psi(r) = \sqrt{\di{L^2 \over{2m}}} \, \di{\int^r_{r_0}} \di{{ \pm dR}\over{R^2 \sqrt{E-U(R)}}} \label{psi}  \enq   

\vv \nin  It is important to realize that, apart the parameters $E$ and $L$,  this angle is a function of the single variable $r$. At any point, the tangent has only two possible orientations, according to the sign in (\ref{psi}). We can suppose that the plus sign and the minus sign are  for $B_a$ and for $B_b$, respectively. Consequently, $B_a$ is a branch where $r$ is increasing from $r_m$ to $r_M$ and $B_b$ a branch where $r$ is decreasing from $r_M$ to $r_m$. It is then clear that any circle with center $O$ and radius $r_1$ such that $r_m \leq r_1 \leq r_M$ intersects $B_a$ at a point $M_a$ and $B_b$ at a point $M_b$. These two points being equidistant from $O$, their corresponding polar angles, as given by (\ref{psi}), are simply opposite (mod. $2 \pi$). Hence, $r(\psi) = r(-\psi)$, which means that ${\bms {OP}}_0$ is an axis of symmetry. The same  kind of reasoning can be applied as well to the limit case where $P_0$ is a turning point. In this case, the branches $B_a$ and $B_b$ are respectively ending and beginning at $P_0$, and both have the same tangent at this point. We thus conclude that every turning point also defines an axis of symmetry. Moreover, if the trajectory 
crosses again an axis of symmetry defined by a double point in another simple point $P_1$ at a polar angle $\theta_0$, the latter must be a turning point. This is so because $r(\theta)$  should be symmetric relatively to the value $\theta_0$, and since $P_1$ is a simple point, we must have $dr/d\theta(\theta_0) =0$. 

\vv \nin To be more precise, let $I_1$, $I_2$ and $I_3$ three successive turning points at $r=r_m$, and $S_1$, $S_2$ the two turning points at $r =r_M$, following after $I_1$ and $I_2$, respectively. Then, the part of the curve between $I_1$ and $I_2$ is symmetric with respect to ${\bms OS_1}$, and the part of the curve between $I_2$ and $I_3$ is symmetric with respect to ${\bms OS_2}$. Moreover, those two parts of the curve are symmetric with respect to ${\bms OI_2}$. As time passes, the same scenario is repeated with other turning points. Then, if the angle $\Delta \theta$ between ${\bms OI_1}$ and ${\bms OS_1}$, Eq. (\ref{deltet}), is not commensurable with $\pi$, the trajectory fills the entire space of the circular crown. and the previous axes of symmetry being in infinite number cannot be distinguished, except maybe one of them, because the polar equation of the trajectory must reflects the said symmetry. The trajectory has a finite number of turning points and corresponding axes of symmetry if and only if $\Delta \theta$ is commensurable with $\pi$. In that case, it is closed. In general, this case arises only for exceptional values of $E$ and $L$. As we know, only the Newtonian and the Hookean potentials produce closed orbits for {\it all} corresponding admissible values of these parameters. 

\vv \nin Turning points also exist for bounded trajectories inside a disk $r \leq r_0$ or unbounded trajectories outside that disk, likely in a limited number for the latter case, for $r=r_0$. So, their polar equation must also reflects some symmetry, a feature that can be used to define a symmetry vector also for such trajectories.

\vvv \vvv
% ==============================================
\nin {\large \bf Appendix B : About the transformation ${\cal T}_A$}

\setcounter{equation}{0}
\renewcommand{\theequation}{\mbox{C.}\arabic{equation}}

\vvv \nin In Special Relativity, all inertial frames are linked by Lorentz transformations These transformations form of a Lie group, the Lorentz group, generated by a Lie algebra  homomorphic to that of the group $SL(2,C)$. With an appropriate choice of their spatial reference axes, two inertial frames ${\cal R}$ and ${\cal R^\prime}$, having a relative velocity ${\bms V} = V {\bms n}$ along an axis defined by the unit vector ${\bms n}$ ($V$ is positive or negative), can always be connected by a Lorentz boost along ${\bms n}$. This means that given an event whose measured data ``time-position" are $(t, {\bms r})$ in ${\cal R}$ and $(t^\prime, {\bms r^\prime})$ in ${\cal R^\prime}$, these datas can be compared by means of the relations 

$$ c t^\prime = c t \cosh \xi + {\bms n} \cdot {\bms r} \sinh \xi  $$ 
$$ {\bms n} \cdot {\bms r^\prime} = {\bms n}\cdot {\bms r} \cosh \xi + ct \sinh \xi $$
$$ {\bms n} \times {\bms r^\prime} =  {\bms r^\prime} - {\bms n} ({\bms n}\cdot {\bms r^\prime}) = {\bms n} \times {\bms r} = {\bms r} - {\bms n} ({\bms n} \cdot {\bms r}) $$
$$ {\rm with}~~ \xi = \tanh^{-1} \left( \di{V \over c} \right) $$

\nin $c$ being the speed of light in vacuum, or 

\beq ~~~~ c t^\prime = c t \cosh \xi + {\bms n} \cdot {\bms r} \sinh \xi ,~~{\bms r^\prime} = {\bms r} + {\bms n} \left( \cosh \xi -1 \right)( {\bms n}\cdot{\bms r}) + {\bms n} \,ct \sinh \xi  \label{boost} \enq

\vv \nin The Lorentz boost, as defined by Eq. (\ref{boost}), applies as well to any 4-vector $Q$ in Minkowski 4-space. Let $Q_0$ and ${\bms Q}$ its time component and spatial components in ${\cal R}$. In ${\cal R^\prime}$ they are $Q^\prime_0$ and ${\bms Q^\prime}$ and we have 

\beq~~~  Q^\prime_0 = Q_0 \cosh \xi + {\bms n} \cdot {\bms Q} \sinh \xi ,\label{boost2} \enq
$${\bms Q^\prime} = {\bms Q} + {\bms n} \left( \cosh \xi -1 \right)( {\bms n}\cdot{\bms Q}) + {\bms n} \,Q_0 \sinh \xi  $$
  
\vv \nin The similarity between Eqs (\ref{trL}), (\ref{trwi}) and Eq. (\ref{boost2}) leads us  to interpret the transformation ${\cal T}_A  $ as a Lorentz boost acting on the fictitious 4-vector $B$ having $L$ as time component and ${\bms B}$ as its spatial part. This 4-vector has the particularity to be lightlike, because its invariant squared norm is zero : 

$$ B^2 = B^2_0 - {\bms B}^2 = L^2 - L^2 = 0 $$

\vv \nin Next, let us define the rotation $R_A : ({\bms U}, {\bms W}, {\bms \ell}) \rightarrow ({\bms U^\prime}, {\bms W^\prime}, {\bms \ell^\prime})$ induced by ${\cal T}_A$. It is easy to obtain the angle of rotation $\gamma$, using the property ${\rm tr} R_A = 1 + 2 \cos \gamma$. We have then 

$$ 1+ 2 \cos \gamma = {\bms U^\prime}\cdot {\bms U} + {\bms W^\prime}\cdot {\bms W} + {\bms \ell^\prime}\cdot {\bms \ell},$$
\nin whence 
$$ \cos \gamma = \di{1\over \rho} \left[ 1 + ({\bms n}\cdot{\bms W})^2 \left(\cosh \xi -1 \right) + {\bms n}\cdot {\bms W} \sinh \xi \right]  $$
\nin with
$$ \rho = \cosh \xi + {\bms n}\cdot{\bms W} \sinh \xi $$

\nin and 

$$ \sin \gamma = \pm \di{{\sqrt{1 - ({\bms n}\cdot{\bms W})^2}} \over \rho} \left[ \sinh \xi + {\bms n}\cdot{\bms W} \left(\cosh \xi -1\right) \right] $$

\vv \nin The components $N_X$, $N_Y$ and $N_Z$, relative to the basis $({\bms U}, {\bms W}, {\bms \ell})$, of the unit vector ${\bms N}$ defining the axis of rotation are subsequently found using the well known transformation formula 

$$ {\bms a^\prime} = {\bms a} \cos \gamma + ({\bms N}\times{\bms a} ) \sin \gamma + (1 - \cos \gamma) ({\bms N}\cdot{\bms a}) {\bms N} $$

\vv \nin We have 

$$ {\bms U^\prime}\cdot {\bms U} = \cos \gamma + (1-\cos \gamma) N^2_X = 1- ({\bms n}\cdot{\bms U})^2 \di{{\left(\cosh \xi -1 \right) }\over \rho},$$
\nin hence
$$N^2_X = \di{{({\bms n}\cdot{\bms \ell})^2}\over{1-({\bms n}\cdot{\bms W})^2}},  $$ 

\vv \nin accounting for the identity $({\bms n}\cdot {\bms U})^2 +({\bms n}\cdot {\bms W})^2+  ({\bms n}\cdot {\bms \ell})^2 = {\bms n}^2 =1$. In the same way, we obtain 

$$ {\bms \ell^\prime}\cdot {\bms \ell} = \cos \gamma + (1-\cos \gamma) N^2_Z = 1- ({\bms n}\cdot{\bms \ell})^2 \di{{\left(\cosh \xi -1 \right) }\over \rho},$$
\nin hence
$$N^2_Z = \di{{({\bms n}\cdot{\bms U})^2}\over{1-({\bms n}\cdot{\bms W})^2}},  $$ 

\vv \nin Since $N^2_X + N^2_Y +N^2_Z =1$, we must have $N_Y=0$. This can be checked as follows. We have 

$$ {\bms W^\prime}\cdot {\bms W} =  \cos \gamma + (1-\cos \gamma) N^2_Y = \di{1\over \rho} \left[ 1+ ({\bms n}\cdot {\bms W})^2 (\cosh \xi-1) + {\bms n}\cdot{\bms W} \sinh \xi \right]= \cos \gamma,$$
\nin hence $N_Y =0$.

\vv \nin Further, using 

$$ {\bms U^\prime}\cdot{\bms W} = N_Z \sin \gamma = - \di{ {\bms n}\cdot{\bms U} \over \rho} \left[\sinh \xi + {\bms n}\cdot {\bms W} (\cosh \xi -1) \right],~~~$$
$$ {\bms U^\prime}\cdot{\bms \ell} = N_Z N_X (1-\cos \gamma) =-( {\bms n}\cdot{\bms \ell})( {\bms n}\cdot {\bms U}) \di{{\cosh \xi-1}\over \rho} $$

\vv \nin we arrive at 

$$ N_X = \pm \di{{{\bms n}\cdot{\bms \ell}}\over \sqrt{1 -({\bms n}\cdot{\bms W})^2}},~~N_Z = \mp \di{{{\bms n}\cdot{\bms U}}\over \sqrt{1 -({\bms n}\cdot{\bms W})^2}} ,$$
\nin and
$$ N_i = \pm \di{1\over \sqrt{1 -({\bms n}\cdot{\bms W})^2}} n_j \left[ \ell_j U_i - U_j \ell_i \right] = \mp \di{1\over \sqrt{1 -({\bms n}\cdot{\bms W})^2}} \, \epsilon_{ijk} n_j W_k $$
$$ = \mp \di{({\bms n}\times{\bms W})_i \over \sqrt{1 -({\bms n}\cdot{\bms W})^2}} $$

%}
\end{document}